\documentstyle[twocolumn,eqsecnum,pra,aps]{revtex}

\newcommand{\eq}[1]{Eq.~(\ref{#1})}
\newcommand{\brak}[1]{\mbox{$ \langle #1  \rangle$}}

\newcommand{\ket}[1]{\mbox{$ | #1 \rangle $}}

\begin{document}

\title{Relativistic Precession of Quantum Elliptical States in the Coulomb
Potential}
\author{Michael G.A. Crawford \\
Laboratory of Atomic and Solid State Physics,\\ Cornell University, Ithaca, NY, 14853-2501}

\maketitle

\begin{abstract}
A special relativistic perturbation to non-relativistic quantum
mechanics is shown to lead to the special relativistic prediction
for the rate of precession for quantum states in the Coulomb potential.
This behavior is illustrated using SO(4) coherent states as examples.
These states are localized on Kepler ellipses and precess in the
presence of a relativistic perturbation.
\end{abstract}

\section{Introduction}

Classical objects which are bound by forces which vary by the inverse
square of distance have elliptical orbits whose orientation does not
change with time.  It is well known that the inclusion of relativistic
effects causes these elliptical orbits to precess.  In the Coulomb
potential, where interactions are electromagnetic, the rate of this
precession is calculated using special relativity.  This was first
accomplished by Sommerfeld~\cite{sommerfeld21} who, for the purpose
of calculating the fine structure of the hydrogen atomic spectrum in
the old quantum theory, calculated the classical rate of precession
due to a special relativistic treatment of the kinetic energy of a
``spinless'' electron the Coulomb potential.  (Sommerfeld's calculation
ignored electron spin in part because it was not described until 1925
by Uhlenbeck and Goudsmit.)

Turning to quantum mechanics, relativistic behavior is here introduced as
a perturbation to the non-relativistic Coulomb potential.  To discuss
this type of precession in quantum mechanics, the first requirement is a
quantum mechanical orbit.  Eigenstates of the unperturbed Hamiltonian
are obvious candidates, but the standard states $\ket{n,\ell,m}$
are certainly not elliptical.  However, for this system there exist
generalized coherent states which are localized along bound classical
trajectories of arbitrary eccentricity~\cite{gay89}.  These states are
formed of the eigenstates pertaining to a single energy level and are
hence themselves eigenstates of the Hamiltonian, stationary in time.
With the introduction of a special relativistic perturbation, the states
will change in time, and in the limit of large quantum number, should
precess at the rate predicted by special relativity.  As shown below,
this is indeed the case.

\section{Estimating the rate of precession}

In the non-relativistic theory, the unperturbed Hamiltonian is
given by
\begin{equation}
\hat{H}_{0} = \frac{\hat{p}^{2}}{2m} - \frac{Ze^{2}}{\hat{r}}
\end{equation}
with energy eigenlevels
\begin{equation}
E^{(0)}_{n} = - m c^{2} \frac{Z^{2} \alpha^{2}}{2 n^{2}}.
\label{dog}
\end{equation}
The perturbation to be considered is given by 
\begin{equation}
\hat{H}_{1} = - \frac{\hat{p}^{4}}{8 m^{3} c^{2}}
\label{filly}
\end{equation}
so that the first order Raleigh-Schr\"odinger perturbative correction to
the energies is given by
\begin{equation}
E^{(1)}_{n,\ell} =
- m c^{2} \frac{Z^{4} \alpha^{4}}{2 n^{3}}
\left ( \frac{1}{\ell + \frac{1}{2}} - \frac{3}{4n} \right ),
\label{ewe}
\end{equation}
in which $m$ is the electron mass,
$c$ is the speed of light,
$Z$ is the atomic number,
$\alpha=e^{2}/\hbar c$ is the fine structure constant,
$e$ is the electron charge,
$n$ is the total quantum number,
and $\ell$ is the quantum number pertaining to angular momentum, the
eigenvalues of $\hat{L}^{2}$ given by $\hbar^{2}\ell(\ell+1)$.
The perturbation \eq{filly} is obtained by expanding the special
relativistic expression for the kinetic energy and retaining the next term
after the Newtonian term.  No spin-orbit coupling is considered
in parallel to the classical theory to which these calculations will be
compared.

The classical period may be extracted from the quantum spectrum
\eq{dog} as follows.  The expansion of the energy
eigenlevels about $n=\brak{n}$ is given by
\begin{eqnarray}
E^{(0)}_{n} &=& - m c^{2} \frac{Z^{2} \alpha^{2}}{2}
\left ( \frac{1}{\brak{n}^{2}}
+ \frac{2}{\brak{n}^{3}} (n-\brak{n}) \right .
\label{gerbil} \\
& & \left . {}
- \frac{3}{\brak{n}^{4}} (n-\brak{n})^{2} + O((n-\brak{n})^{3}) \right ).
\nonumber
\end{eqnarray}
Now assume that the
system is in some state $\ket{\psi}$ with average total quantum
number $\brak{n}$ and uncertainty $(\Delta n)^{2} = \brak{n^{2}}-\brak{n}^2$.
If the quantum time evolution is expressed in terms
of the eigenstate expansion,
\begin{equation}
\ket{\psi(t)} = \sum_{n} e^{-i E_{n} t/\hbar} c_{n} \ket{n},
\end{equation}
the first term in \eq{gerbil} leads to an overall phase factor which
may be dropped, and the second term will lead to phase factors that are
integer multiples of $2\pi$ when
\begin{equation}
t = T_{cl} = \frac{2 \pi \hbar^{3} \brak{n}^{3}}{m Z^{2} e^{4}}.
\end{equation}
This gives the classical period of a classical trajectory with
energy $E_{n}^{(0)}$ exactly, a circumstance closely connected to
the success of Bohr-Sommerfeld quantization in hydrogenic atoms.  Also,
for hydrogenic wave functions, radial expectation values are given by
\begin{equation}
r_{n} = \brak{n,\ell,m|\hat{r}|n,\ell,m} = \frac{n^{2} \hbar^{2}}{Z m e^{2}}.
\label{colt}
\end{equation}
Substitution of this value into $T_{cl}$ yields
\begin{equation}
T_{cl}^{2} = \frac{4 \pi^{2} r_{n}^{3}}{Z^{4} e^{4}},
\end{equation}
a quantum rendering of Kepler's third law.

If the expansion \eq{gerbil} terminated with the linear term, then
periodic motion would continue indefinitely.  However, the
quadratic and higher terms contribute a dephasing influence and the
approximation of periodic behaviour eventually breaks down.  To
estimate the size of this dephasing influence, 
evaluate the quadratic term at the edge
of the distribution in $n$ ($n=\brak{n} \pm \Delta n$) at $t=T_{cl}$
to yield the ``test quantity''
\begin{equation}
\delta \phi = 3 \pi \frac{(\Delta n)^{2}}{\brak{n}}.
\label{lemur}
\end{equation}
Hence, in the special case of a state localized in position and
momentum, Ehrenfest's equations indicate an initial trajectory
which follows a classical trajectory.  In the additional case that
$\delta \phi \ll 2 \pi$, the state achieves at least a full period
of behavior approximating classical behavior.  These considerations
are similar to those used employed in describing wave function
revivals~\cite{averbukh89,nauenberg90,yeazell90,crawford00}.

Now treating the relativistic perturbation to kinetic energy \eq{ewe}
in a similar manner, the expansion about $\ell=\brak{\ell}$ of the
perturbation \eq{ewe} is given by
\begin{eqnarray}
E^{(1)}_{n,\ell} &=& 
m c^{2} \frac{Z^{4} \alpha^{4}}{2 n^{3}}
\left [
- \left ( \frac{1}{\brak{\ell} + \frac{1}{2}} - \frac{3}{4n} \right )
+ \frac{(\ell - \brak{\ell})}{(\brak{\ell} + \frac{1}{2})^{2}} \right .
\nonumber \\
& & \left . {}
- \frac{(\ell - \brak{\ell})^{2}}{(\brak{\ell} + \frac{1}{2})^{3}}
+ O((\ell - \brak{\ell})^{3}) \right ].
\label{cat}
\end{eqnarray}
Suppose the system is now in a state composed of eigenstates pertaining
to the same $n$, degenerate in the unperturbed system.
Then, the term independent
of $(\ell-\brak{\ell})$ in \eq{cat} leads to overall phase factors,
and the linear term leads to integer multiples of $2\pi$ when
\begin{equation}
t=T_{p} = \frac{4 \pi \hbar n^{3}}{m c^{2} Z^{4} \alpha^{4}}
\left (\brak{\ell} + \frac{1}{2} \right )^{2}.
\end{equation}
This may be written in terms of $T_{cl}$, yielding
\begin{equation}
T_{p} = \frac{2 T_{cl}}{Z^{2} \alpha^{2}}
\left (\brak{\ell} + \frac{1}{2} \right )^{2}.
\label{pig}
\end{equation}
Note that $T_{p} \gg T_{cl}$ which follows from $E^{(1)}_{n,\ell} \ll
E^{(0)}_{n}$.

With the additional assumption that the system is in a state built
up upon a particular classical orbit (such states are described in
the next section), the evolution which occurs on a time scale $T_{p}$
derived above must be of the same nature as classical precession due to
relativity, that is, in-plane rotation of the state about the origin with
a period of $T_{p}$.  If this is the case, then the angle of rotation
per classical period $T_{cl}$ is given by
\begin{equation}
\delta \omega = \frac{\pi Z^{2} \alpha^{2}}
{(\brak{\ell} + \frac{1}{2})^{2}}.
\label{husky}
\end{equation}

Continuing the analogous arguments which lead
to \eq{lemur}, the quadratic term of \eq{cat} evaluated at $t=T_{p}$
at the edges of the distribution in $\ell$ yields the test quantity
\begin{equation}
\delta \phi = \frac{m c^{2} Z^{4} \alpha^{4} (\Delta \ell)^{2}}
{2 n^{3} (\brak{\ell}+\frac{1}{2})^{3}}
\frac{T_{p}}{\hbar} =
\frac{2 \pi (\Delta \ell)^{2}}
{\brak{\ell}+\frac{1}{2}}.
\label{snake}
\end{equation}
Again, if $\delta \phi$ is small, then the state at $t=T_{p}$ will
approximately resemble the state at $t=0$.  This condition is best
satisfied by Rydberg states with large angular momentum.

From Bergmann~\cite{bergmann42}, the classical rate of perihelion
precession according to special relativity is given by\footnote{This
rate is never observed in nature since those objects whose orbits
are observed to precess are invariably gravitationally bound, requiring
a general relativistic treatment.  Such a treatment yields a rate,
observed in the case of Mercury\cite{will93b}, of
$\delta \omega = 6 \pi G^{2} M^{2} m^{2} / c^{2} L^{2}$.}
\begin{equation}
\delta \omega = \pi \frac{G^2 M^2 m^2}{c^{2} L^{2}},
\label{hog}
\end{equation}
measured as a change in angle per classical period of the vector
pointing from perihelion to aphelion.  In this expression, $G$ is
the gravitational constant, $M$ is the (large) mass of the source
of the gravitational field, $m$ is the (small) mass of body in
orbit, and $L$ is the orbital angular momentum of the orbiting
body.
To compare the quantum prediction \eq{husky} with the classical
rate \eq{hog}, replace
the strength of the Coulomb potential ($Z e^{2}$) with that of the
Kepler problem ($GMm$), and note that the total angular momentum
squared will be close to $\hbar^{2} (\brak{\ell}+\frac{1}{2})^{2}$
given the above assumptions that $\brak{\ell}$ is large and $(\Delta
\ell)^{2}$ is small:
\begin{equation}
\brak{\ell(\ell+1)} - (\brak{\ell}+\frac{1}{2})^{2} =
(\Delta \ell)^{2} - \frac{1}{4}.
\end{equation}
These substitutions render the quantum estimate \eq{husky} identical to
\eq{hog} as anticipated.

\section{SO(4) coherent states}

The generalization of coherent states due to
Perelomov~\cite{perelomov72,perelomov86} is useful in this context.
This generalization rests upon the group structure of the system
in question.  In the case of the hydrogen atom, the dynamical group
is SO(4,2)~\cite{barut71a,paldus96,adams88}.  For the present
purposes, it is not necessary to engage the entire group; the
degeneracy group SO(4) contains sufficient structure.  Treatments
of coherent states of this description are existent in the
literature~\cite{gay89,crawford00}.

The realization of SO(4) which describes the degeneracy of the
hydrogen atom spectrum is given by the elements of the angular
momentum operator and those of the scaled Laplace-Runge-Lenz vector.
Given a classical orbit, the classical version of the Laplace-Runge-Lenz
vector is proportional in magnitude to the eccentricity and is
aligned parallel to the major axis.  The corresponding quantum
operators in atomic units are given by~\cite{adams88}
\begin{eqnarray}
\hat{\mathbf{L}} &=& \hat{\mathbf{r}} \times \hat{\mathbf{p}},\\
\hat{\mathbf{A}} &=& \frac{1}{2} \hat{\mathbf{r}} \hat{p}^{2}
- \hat{\mathbf{p}} (\hat{\mathbf{r}} \cdot \hat{\mathbf{p}})
- \frac{1}{2}\hat{\mathbf{r}}.
\end{eqnarray}
All of these operators commute with the Hamiltonian, leading to
conservation of these quantities under time evolution.  Conservation
of angular momentum is to be expected in this (spherically symmetric)
potential, but conservation of the $\hat{\mathbf{A}}$ (a simple derivation
of which is given by Wulfman~\cite{wulfman71}) is a unique property of
the non-relativistic Coulomb potential.  The classical interpretation
of the invariance of the Laplace-Runge-Lenz vector is that elliptical
orbits do not precess.

These operators satisfy the commutation relations
\begin{equation}
[\hat{L}_{j}, \hat{L}_{k}] = i \epsilon_{jkl} \hat{L}_{l}, \quad
[\hat{A}_{j}, \hat{A}_{k}] = i \epsilon_{jkl} \hat{L}_{l}, \quad
[\hat{L}_{j}, \hat{A}_{k}] = i \epsilon_{jkl} \hat{A}_{l}.
\end{equation}
These six operators may be decoupled into two groups of three operators via
\begin{equation}
\hat{M}_{j} = \frac{1}{2} (\hat{L}_{j} + \hat{A}_{j}), \quad
\hat{N}_{j} = \frac{1}{2} (\hat{L}_{j} - \hat{A}_{j}),
\end{equation}
which commute according to
\begin{equation}
{}[\hat{M}_{j}, \hat{M}_{k}] = i \epsilon_{jkl} \hat{M}_{l}, \quad
{}[\hat{N}_{j}, \hat{N}_{k}] = i \epsilon_{jkl} \hat{N}_{l}, \quad
{}[\hat{M}_{j}, \hat{N}_{k}] = 0.
\end{equation}
In terms of these new operators, it is clear that
$\mbox{SO}(4) = \mbox{SO}(3) \otimes \mbox{SO}(3)$.

The group SO(4) has two Casimir operators, given by
\begin{eqnarray}
\hat{C}_{1} &=& \hat{L}^{2} + \hat{A}^{2} = 2(\hat{M}^{2} + \hat{N}^{2}), \\
\hat{C}_{2} &=& \hat{\mathbf{L}} \cdot \hat{\mathbf{A}}
= \hat{M}^{2} - \hat{N}^{2} .
\end{eqnarray}
In the hydrogenic realization of this group, $\hat{C}_{2}=0$ so
that quantum mechanically and classically, angular momentum is
perpendicular to the Laplace-Runge-Lenz vector.  The second Casimir
operator also indicates that the dimensions of the irreducible
representations of the SO(3) generated by the $\hat{M}_{k}$ and by
the $\hat{N}_{k}$ are equal (say, to $n=2j+1$) so that the dimensions
of the relevant unitary irreducible representations of SO(4) are
$n^{2}$, the famous degeneracy of the hydrogen atom energy spectrum.
Thus the first Casimir operator is equal to
\begin{equation}
\hat{C}_{1} = 4j(j+1) = n^{2}-1,
\end{equation}
representing a constraint on the sum $\brak{\hat{L}^{2} + \hat{A}^{2}}$.

Turning to the SO(4) coherent states, Since $\mbox{SO}(4) =
\mbox{SO}(3) \otimes \mbox{SO}(3)$, the SO(4) coherent states may
be expressed as the direct product of two SO(3) coherent states
which are themselves standard in the
literature~\cite{klauder85,perelomov86,feng90}.  The SO(3) coherent
states are given by
\begin{equation}
\ket{j,\zeta} = \sum_{m=-j}^{j}
\left ( \frac{(2j)!}{(j+m)!(j-m)!} \right ) ^{1/2}
\frac{\zeta^{j+m}}{(1+|\zeta|^{2})^{j}} \ket{j,m},
\label{mouse}
\end{equation}
parameterized by the complex valued $\zeta$.  In these states,
expectation values of the angular momentum operators are given by
\begin{equation}
\brak{\hat{\mathbf{J}}} =  \frac{2j}{1+|\zeta|^{2}}
\left ( \mbox{Re}(\zeta), - \mbox{Im}(\zeta),
\frac{1}{2} (|\zeta|^{2}-1) \right ),
\label{cow}
\end{equation}
with $\hat{\mathbf{J}}$ standing for $\hat{\mathbf{M}}$ or $\hat{\mathbf{N}}$
as the case may be.  With these expressions in mind, the SO(4) coherent
states are given by
\begin{equation}
\ket{n,\zeta_{1},\zeta_{2}} = \ket{j,\zeta_{1}}\ket{j,\zeta_{2}}.
\label{ram}
\end{equation}
Here, $n=2j+1$, and $\zeta_1$ and $\zeta_{2}$ parameterize the SO(3)
coherent states pertaining to $\hat{\mathbf{M}}$ and $\hat{\mathbf{N}}$
respectively.  The expectation values of the angular momentum and
Laplace-Runge-Lenz operators may be regained through \eq{cow} and the
relations $\hat{\mathbf{L}} = \hat{\mathbf{M}} + \hat{\mathbf{N}}$
and $\hat{\mathbf{A}} = \hat{\mathbf{M}} - \hat{\mathbf{N}}$. (The
substitution of \eq{mouse} into \eq{ram} yields a double sum over the
direct product states $\ket{j,m_{1}}\ket{j,m_{2}}$.  These states may be
calculated by relating them to the standard hydrogenic eigenstates
$\ket{n,\ell,m}$ via the Clebsch-Gordon coefficients.)

For the purpose of visualization, spherical symmetry permits setting
$\brak{\hat{\mathbf{L}}}$ parallel to the $z$-axis, achieved by setting
$\zeta_{2} = - \zeta_{1}$.  Since $\hat{\mathbf{L}} \cdot \hat{\mathbf{A}}
= 0$, the vector $\brak{\hat{\mathbf{A}}}$ may be oriented parallel to
the $x$-axis accomplished by setting the imaginary parts of $\zeta_{1}$
and $\zeta_{2}$ equal to zero.  These identifications reduce the
problem to the variation of a single real parameter, say $\eta =
\mbox{Re}(\zeta_{1})$.  In terms of this parameter, expectation values
are given by
\begin{equation}
\brak{\hat{L}_{3}} = \frac{2j(\eta^{2}-1)}{1+\eta^{2}}, \quad
\brak{\hat{A}_{1}} = \frac{4j\eta}{1+\eta^{2}}.
\end{equation}
With the magnitude of the classical Laplace-Runge-Lenz vector being
proportional to the eccentricity $\epsilon$ of the orbit, the
quantum calculation leads to
\begin{equation}
\epsilon = \frac{2 \eta}{1+\eta^{2}}.
\label{bull}
\end{equation}
Also in terms of $\eta$, the total angular momentum may be expressed as
\begin{equation}
\brak{\hat{L}^{2}} = 2j(j+1) + 2j^{2} \frac{\eta^{4}-6\eta^{2}+1}
{(1+\eta^{2})^{2}}.
\end{equation}
This expression follows from 
$\brak{\hat{L}^{2}} = \brak{(\hat{\mathbf{M}}+\hat{\mathbf{N}})
\cdot (\hat{\mathbf{M}}+\hat{\mathbf{N}}) }$, \eq{cow}, and
$\brak{\hat{\mathbf{M}} \cdot \hat{\mathbf{N}}} = 
\brak{\hat{\mathbf{M}}} \cdot \brak{\hat{\mathbf{N}}}$, the last
of which since $[\hat{M}_{j},\hat{N}_{k}]=0$.

\begin{figure}[h!]
\begin{picture}(0,0)%
\includegraphics{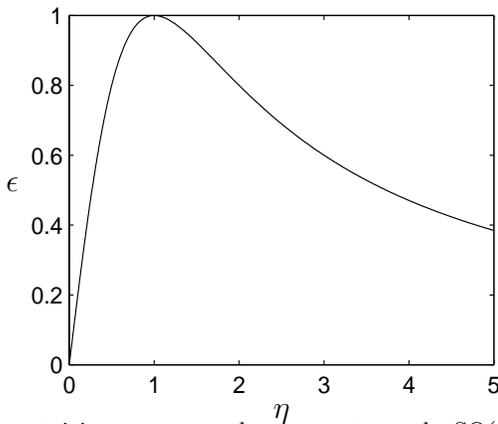}%
\end{picture}%
\setlength{\unitlength}{3947sp}%
\begingroup\makeatletter\ifx\SetFigFont\undefined%
\gdef\SetFigFont#1#2#3#4#5{%
  \reset@font\fontsize{#1}{#2pt}%
  \fontfamily{#3}\fontseries{#4}\fontshape{#5}%
  \selectfont}%
\fi\endgroup%
\begin{picture}(3872,2674)(41,-2086)
\put(826,-661){\makebox(0,0)[rb]{\smash{\SetFigFont{12}{14.4}{\rmdefault}{\mddefault}{\updefault}$\epsilon$}}}
\put(2476,-2086){\makebox(0,0)[b]{\smash{\SetFigFont{12}{14.4}{\familydefault}{\mddefault}{\updefault}$\eta$}}}
\end{picture}
\caption{Eccentricity $\epsilon$ versus $\eta$, the parameter to the
SO(4) coherent state.}
\label{epsilon}
\end{figure}

By virtue of their construction, the SO(4) coherent states are minimum
uncertainty states of rotated 4-dimensional angular momentum operators.
One therefore expects that for moderate 4-dimensional rotations,
the states will remain localized in angular momentum and in the
Laplace-Runge-Lenz vector, that is, localized about a particular
classical orbit.  Hence, given a value of $\eta$, the approximate
geometry is given by \eq{bull}, with a semi-major axis given by
\eq{colt}.

\section{Precessing coherent states}

\begin{figure}[h!]
\begin{picture}(0,0)%
\includegraphics{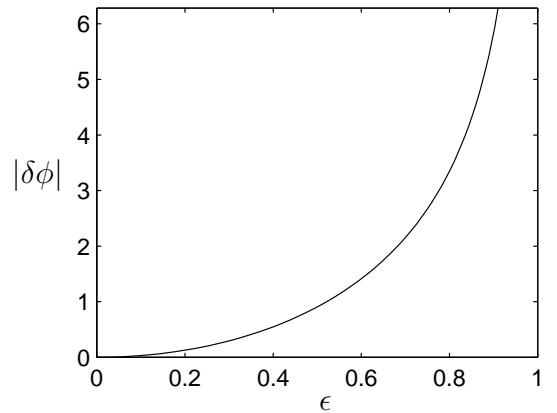}%
\end{picture}%
\setlength{\unitlength}{3947sp}%
\begingroup\makeatletter\ifx\SetFigFont\undefined%
\gdef\SetFigFont#1#2#3#4#5{%
  \reset@font\fontsize{#1}{#2pt}%
  \fontfamily{#3}\fontseries{#4}\fontshape{#5}%
  \selectfont}%
\fi\endgroup%
\begin{picture}(4117,2732)(-40,-2144)
\put(2640,-2086){\makebox(0,0)[b]{\smash{\SetFigFont{12}{14.4}{\familydefault}{\mddefault}{\updefault}$\epsilon$}}}
\put(990,-661){\makebox(0,0)[rb]{\smash{\SetFigFont{12}{14.4}{\rmdefault}{\mddefault}{\updefault}$|\delta \phi|$}}}
\end{picture}
\caption{Test parameter $\delta \phi$ versus eccentricity $\epsilon$.}
\label{phi}
\end{figure}

The value of the test quantity $\delta \phi$ given by \eq{snake}
is expressible in terms of the SO(4) coherent state parameter
$\eta$.  With expectation values taken in an SO(3) coherent state,
\begin{equation}
(\Delta \hat{J}_{3})^{2} = 2 j \frac{\eta^{2}}{(1+\eta^{2})^{2}},
\end{equation}
and
\begin{equation}
(\Delta \ell)^{2} = (\Delta \hat{L}_{3})^{2} = (\Delta \hat{M}_{3})^{2} +
(\Delta \hat{N}_{3})^{2},
\end{equation}
since $\hat{M}_{j}$ and $\hat{N}_{j}$ commute.  With these in mind, and
dropping the $\frac{1}{2}$ from the denominator of $\delta \phi$ since
$\brak{\ell}$ is large,
\begin{equation}
\delta \phi = 2 \pi \frac{\eta^{2}}{\eta^{4}-1}.
\end{equation}
Therefore, in the sense that $\delta \phi = 0$, the precession is best
for $\eta=0$ or as $\eta
\rightarrow \infty$.  In both cases, though, from \eq{bull}, the resultant
orbits will be circular and no precession can be observed:  There is
a certain tradeoff between the observability and ``coherence'' of the
precession measured by $\delta \phi$.  In fact, it is convenient
to express $\delta \phi$ in terms of $\epsilon$ from \eq{bull} so as to
connect this test with a more physical or geometrical quantity:
\begin{equation}
\delta \phi = \frac{ 2 \pi \epsilon^{2}
(2 - \epsilon^{2} \pm 2  \sqrt{1 - \epsilon^{2}}) }
{ 8-8\epsilon^{2} \pm 4(2-\epsilon^{2})\sqrt{1-\epsilon^{2}}}
=\frac{\pi}{2} \epsilon^{2} + O(\epsilon^{4}),
\end{equation}
where the positive sign is appropriate for $\eta>1$ and the negative for
$\eta<1$.  As shown in Figure~\ref{phi}, as the eccentricity increases,
the degree to which the wave functions remain assembled decreases.

\begin{figure}[h!]
\begin{picture}(0,0)%
\includegraphics{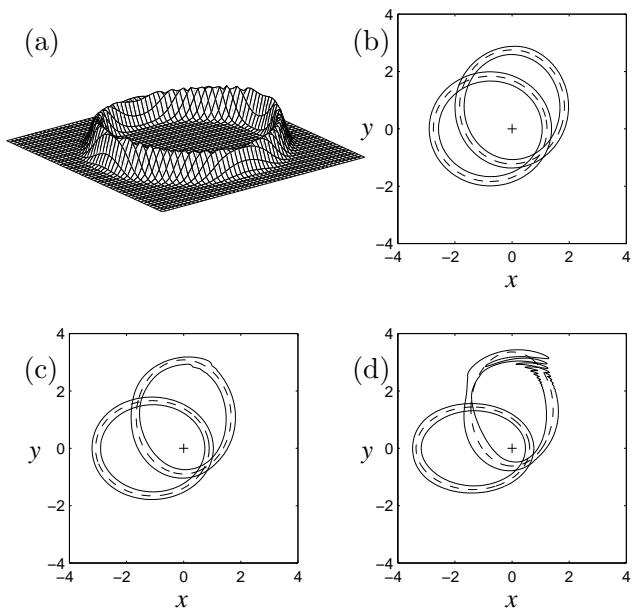}%
\end{picture}%
\setlength{\unitlength}{2171sp}%
\begingroup\makeatletter\ifx\SetFigFont\undefined%
\gdef\SetFigFont#1#2#3#4#5{%
  \reset@font\fontsize{#1}{#2pt}%
  \fontfamily{#3}\fontseries{#4}\fontshape{#5}%
  \selectfont}%
\fi\endgroup%
\begin{picture}(7228,7903)(664,-8477)
\put(4651,-5761){\makebox(0,0)[lb]{\smash{\SetFigFont{11}{13.2}{\familydefault}{\mddefault}{\updefault}(d)}}}
\put(901,-5761){\makebox(0,0)[lb]{\smash{\SetFigFont{11}{13.2}{\familydefault}{\mddefault}{\updefault}(c)}}}
\put(4651,-2011){\makebox(0,0)[lb]{\smash{\SetFigFont{11}{13.2}{\familydefault}{\mddefault}{\updefault}(b)}}}
\put(901,-2011){\makebox(0,0)[lb]{\smash{\SetFigFont{11}{13.2}{\familydefault}{\mddefault}{\updefault}(a)}}}
\end{picture}
\caption{Some examples of precessing SO(4) coherent states: (a) In the
141st energy level, at
$t=0$, with $\eta = 0.2$ (or $\epsilon=0.385$) on the $x$-$y$ plane
with amplitude proportional to
$|\brak{\mathbf{r}|141,\zeta_{1},\zeta_{2}}|^{2}$.
(b) The same state as (a), with
position plotted in $10^{4}$ atomic units, and the origin
located at the + symbol.
The state at $t=0$ is oriented
horizontally, and $t=\frac{1}{4}T_{p}$ oriented vertically.  The solid
lines depict a single contour on the quantum wave function, and the dashed
lines depict the classical orbit precessed according to the special
relativistic prediction.
(c) The same as (b), but with $\eta=0.3$ (or $\epsilon=0.550$).
(d) The same as (b), but with $\eta=0.4$ (or $\epsilon=0.690$). }
\label{picture}
\end{figure}

For some examples of precessing coherent states, examine Figure
\ref{picture}.  This figure depicts a mesh plot of an SO(4) coherent
state at $t=0$, and a sequence of overlayed contour plots showing
states precessed from $t=0$ to $t=\frac{1}{4}T_{p}$.  To appreciate
the physical scale of these simulations, the field of view in all
cases is 4.23 $\mu$m across and with states in the 141st energy
level, the classical period is $4.25\times 10^{-10}$ seconds.  In
Figure \ref{picture}(b), with $\epsilon=0.385$, the precession
period is 0.266 seconds or $6.26\times 10^{8}$ classical periods,
which on atomic scales, is a very long time.  In the subsequent
images, the rate of precession is larger owing to the larger
eccentricity so that the precession period of the state depicted
in Figure \ref{picture}(d) is 0.164 seconds.  This interval is still
much longer than the longest times Rydberg states are observed in
experimental setups, typically 1 nanosecond\cite{yeazell90}.

As apparent from the images, the quantum rate of precession agrees
with the classical special relativistic prediction.  In the case of Figure
\ref{picture}(b), though only shown until $t=\frac{1}{4} T_{p}$,  the
state remains localized on the ellipse up to the full period of the
precession.  Not surprisingly, for increased eccentricities, the degree of
``coherence'' of the state decreases, leading to the decay of the state
as depicted in Figure \ref{picture}(d) after only a quarter precession
period.

\section{Conclusions}

The perturbation \eq{filly} was chosen so that a direct comparison could
be made between the quantum mechanical calculation and the special
relativistic calculation due to Sommerfeld.  Therefore, spin effects
were ignored which are equal in size to the kinetic energy perturbation.
This means that the quantities calculated here are not predictions to be
tested in the laboratory.  The purpose has been to show that a simple
analysis of a quantum perturbation can reproduce the results of a more
involved classical analysis.  In particular, a special relativistic
perturbation to non-relativistic quantum mechanics leads to an agreement
with classical special relativity in a large quantum number limit.
A similar notion of agreement may be found in the work of McRae and
Vrscay~\cite{mcrae97} who have studied correspondence between quantum and
classical perturbation schemes.  In their work, as well as in the present
paper, it transpires that the simplest route to the determination of a
classical perturbation may be through the classical limit of a quantum
perturbation.

\section{Acknowledgments}

This work was supported by the Natural Sciences and Engineering Research
Council of Canada, and the National Science Foundation.  The author would
also like to acknowledge useful discussions with Prof. N.W. Ashcroft of
the Department of Physics, Cornell University, and Profs. E.R. Vrscay and
J. Paldus of the Department of Applied Mathematics, University of Waterloo.


\begin{thebibliography}{10}

\bibitem{sommerfeld21}
A.~Sommerfeld.
\newblock {\em Atombau und Spektrallinien}.
\newblock F. Vieweg und Sohn, Braunschweig, Germany, 1921.

\bibitem{gay89}
J.C. Gay, D.~Delande, and A.~Bommier.
\newblock Atomic quantum states with maximum localization of classical
  elliptical orbits.
\newblock {\em Phys. Rev. A}, 39(12):6587--6590, 1989.

\bibitem{averbukh89}
I.S. Averbukh and N.F. Perelman.
\newblock Fractional revivals: Universality in the long-term evolution of
  quantum wave packets beyond the correspondence principle dynamics.
\newblock {\em Phys. Lett. A}, 139(9):449--453, 1989.

\bibitem{nauenberg90}
M.~Nauenberg.
\newblock Autocorrelation function and quantum recurrence of wavepackets.
\newblock {\em J. Phys. B}, 23:L385--L390, 1990.

\bibitem{yeazell90}
J.A. Yeazell, M.~Mallalieu, and Jr. C.R.~Stroud.
\newblock Observation of the collapse and revival of a {Rydberg} electronic
  wave packet.
\newblock {\em Phys. Rev. Lett.}, 64(17):2007--2010, 1990.

\bibitem{crawford00}
M.G.A. Crawford.
\newblock Temporally stable coherent states in energy degenerate systems: The
  hydrogen atom.
\newblock {\em Phys. Rev. A}, 62(1):012104--1--7, 2000.

\bibitem{bergmann42}
G.~Bergmann.
\newblock {\em Introduction to the Theory of Relativity}.
\newblock Prentice-Hall, Englewood Cliffs, NJ, 1942.

\bibitem{will93b}
C.M. Will.
\newblock {\em Theory and Experiment in Gravitational Physics}.
\newblock Cambridge University Press, Cambridge, revised edition, 1993.

\bibitem{perelomov72}
A.M. Perelomov.
\newblock Coherent states for arbitrary {Lie} groups.
\newblock {\em Commun. Math. Phys.}, 26:222--236, 1972.

\bibitem{perelomov86}
A.M. Perelomov.
\newblock {\em Generalized Coherent States and Their Applications}.
\newblock Springer-Verlag, London, 1986.

\bibitem{barut71a}
A.O. Barut and G.L. Bornzin.
\newblock {SO}(4,2)-{Formulation} of the symmetry braking in relativistic
  {Kepler} problems with or without magnetic charges.
\newblock {\em J. Math. Phys.}, 12(5):841--846, 1971.

\bibitem{paldus96}
J.~Paldus.
\newblock Dynamical groups.
\newblock In G.W.F. Drake, editor, {\em Atomic, Molecular and Optical Physics
  Handbook}. AIP Press, Woodbury, New York, 1996.

\bibitem{adams88}
B.G. Adams, J.~\v{C}\'\i\v{z}ek, and J.~Paldus.
\newblock Lie algebraic methods and their applications to simple quantum
  systems.
\newblock In {\em Advances in Quantum Chemistry, Vol 19}. Academic Press, New
  York, 1988.

\bibitem{wulfman71}
C.E. Wulfman.
\newblock Dynamical groups in atomic and molecular physics.
\newblock In E.M. Loebl, editor, {\em Group Theory and Its Applications},
  volume~II. Academic Press, New York, 1971.

\bibitem{klauder85}
J.R. Klauder and B.-S. Skagerstam, editors.
\newblock {\em Coherent States: Applications in Physics and Mathematical
  Physics}.
\newblock World Scientific, Singapore, 1985.

\bibitem{feng90}
W.-M. Zhang, D.H. Feng, and R.~Gilmore.
\newblock Coherent states: Theory and some applications.
\newblock {\em Rev. Mod. Phys.}, 62(4):867--927, 1990.

\bibitem{mcrae97}
S.M. McRae and E.R. Vrscay.
\newblock Perturbation theory and the classical limit of quantum mechanics.
\newblock {\em J. Math. Phys.}, 38(6):2899, 1997.

\end{thebibliography}

\end{document}